\documentclass{article}

\usepackage{amsthm,amsmath,amsfonts,amssymb}
\usepackage[numbers]{natbib}
\usepackage[utf8]{inputenc} 
\usepackage{graphicx}
\usepackage{bm}
\usepackage{url}
\usepackage{authblk}
\usepackage{csquotes}
\usepackage{booktabs}
\usepackage{subcaption}
\usepackage[linesnumbered, ruled]{algorithm2e} 
\usepackage[colorlinks=false,hidelinks]{hyperref}
\usepackage[noabbrev,capitalize,nameinlink]{cleveref}

\newcommand{\etal}{\textit{et. al.}}
\newcommand{\bfX}{\mathbf{X}}
\newcommand{\bfXT}{\bfX^{\intercal}}
\newcommand{\bfx}{\mathbf{x}}
\newcommand{\bfg}{\mathbf{g}}
\newcommand{\bfH}{\mathbf{H}}
\newcommand{\bfS}{\mathbf{S}}
\newcommand{\bfy}{\mathbf{y}}
\newcommand{\bfbeta}{\boldsymbol{\beta}}
\newcommand{\bfp}{\mathbf{p}}
\newcommand{\bfw}{\mathbf{w}}
\newcommand{\bfW}{\mathbf{W}}
\newcommand{\bfz}{\mathbf{z}}
\newcommand{\bfs}{\mathbf{s}}
\newcommand{\bfM}{\mathbf{M}}
\newcommand{\bfb}{\mathbf{b}}

\newcommand{\bferr}{\mathbf{err}}
\newcommand{\bfnum}{\mathbf{numerator}}
\newcommand{\bfden}{\mathbf{denominator}}
\newcommand{\POtwo}[1]{\lceil #1 \rceil_{\text{PO}2}}
\newcommand{\vbrak}[1]{vector$\langle$#1$\rangle$}
\newcommand{\aref}[1]{Algorithm~\ref{#1}}
\newcommand{\tref}[1]{Table~\ref{#1}}
\newcommand{\fref}[1]{Figure~\ref{#1}}

\SetKwInput{Initialization}{Initialization}

\begin{document}

\title{Achieving GWAS with Homomorphic Encryption}


\author[1]{Jun Jie Sim}
\author[1]{Fook Mun Chan}
\author[1]{Shibin Chen}
\author[1]{Benjamin Hong Meng Tan}
\author[1]{Khin Mi Mi Aung}

\affil[1]{Institute of Infocomm Research, Singapore}

\maketitle

\begin{abstract} 
One way of investigating how genes affect human traits would be with a genome-wide association study (GWAS).
Genetic markers, known as single-nucleotide polymorphism (SNP), are used in GWAS.
This raises privacy and security concerns as these genetic markers can be used to identify individuals uniquely.
This problem is further exacerbated by a large number of SNPs needed, which produce reliable results at a higher risk of compromising the privacy of participants.

We describe a method using homomorphic encryption (HE) to perform GWAS in a secure and private setting.
This work is based on a proposed  algorithm.
Our solution mainly involves homomorphically encrypted matrix operations and suitable approximations that adapts the semi-parallel GWAS algorithm for HE.
We leverage upon the complex space of the CKKS encryption scheme to increase the number of SNPs that can be packed within a ciphertext.
We have also developed a cache module that manages ciphertexts, reducing the memory footprint.

We have implemented our solution over two HE open source libraries, HEAAN and SEAL.
Our best implementation took $24.70$ minutes for a dataset with $245$ samples, over $4$ covariates and $10643$ SNPs.

We demonstrate that it is possible to achieve GWAS with homomorphic encryption with suitable approximations.
\end{abstract}

\section*{Background}
Genome-wide association study (GWAS) compares genetic variants, single-nucleotide polymorphisms (SNP), to see if these variants are associated with a particular trait.
The model used in GWAS is essentially logistic regression, evaluated one SNP at a time, corrected with covariates like age, height and weight.
The number of SNPs analyzed can easily grow up to $30$ million.
It is estimated that it can take around $6$ hours for $6000$ samples and $2.5$ million SNPs \cite{GRIMP09}.

Some suggest that cloud computing could offer a cost-effective and scalable alternative that allows research to be done, given the exponential growth of genomic data and increasing computational complexity of genomic data analysis.
However, privacy and security are primary concerns when considering these cloud-based solutions.

It was shown in $2004$ by Lin \etal~\cite{Lin183} that as little as $30$ to $80$ SNPs could identify an individual uniquely.
Homer \etal~\cite{Homer08} further demonstrated that even when DNA samples are mixed among $1000$ other samples, individuals could be identified.
In light of these discoveries, regulations concerning biological data are being updated \cite{Rousseaur17}.
The privacy and security of DNA-related data are now more important than ever.

Homomorphic Encryption (HE) is a form of encryption where functions, $f$, can be evaluated on encrypted data $x_1, \ldots, x_n$, yielding ciphertexts that decrypt to $f(x_1, \ldots, x_n)$.
Putting it in the context of GWAS, genomic data can be homomorphically encrypted and sent to a computational server.
The server then performs the GWAS computations on the encrypted data, before sending the encrypted outcome to the data owner for decryption.
We argue that this would ensure the privacy and security of genomic data:
Throughout the entire process, there is no instance where the server can access the data in its raw, unencrypted form, preserving the privacy of the data.
Additionally, since the data is encrypted, no adversary would be able to make sense of the ciphertexts.
The data is thus secured on the computational server.

Motivated by these concerns, the iDASH Privacy \& Security Workshop \cite{iDash} has organized several competitions on secure genomics analysis since $2011$.
The aim of these competitions is to evaluate methods that provide data confidentiality during analysis in a cloud environment.

In this work, we provide a solution to Track $2$ of the iDASH 2018 competition -- \textit{Secure Parallel Genome-Wide Association Studies using Homomorphic Encryption}.
The challenge of this task was to implement the semi-parallel GWAS algorithm proposed by Sikorska \etal~\cite{Sikorska2013}, which outperforms prior methods by about $24$ times, with HE.
This task seeks to advance the practical boundaries of HE, a continuation from last year's HE task which was to implement logistic regression with HE.

We propose a modification of the algorithm by Sikorska \etal~\cite{Sikorska2013} for homomorphically encrypted matrices.
We developed a caching system to minimize memory utilization while maximizing the use of available computational resources.
Our solution also leverages on the complex space of the CKKS encoding to store the SNP matrix and this halved the computation time needed by doubling the number of SNPs processed each time.

Within the constraints of the competition, including a virtual machine with 16GB of memory and 200GB disk space, a security level of at least 128 bits and at most 24 hours of runtime, our solution reported a total computation time of 717.20 minutes.
Our best implementation using a more efficient HE scheme, which was not available during the competition, achieved a runtime of $24.70$ minutes.

In the following section, we will first define some notations used in this paper.
We will begin by describing the CKKS homomorphic encryption scheme that was used to implement the GWAS algorithm.
We describe our methods for manipulating homomorphic matrices that are crucial to our solution.
We start with our implementation of logistic regression with HE.
Following that, we adapted the GWAS algorithm using suitable approximations to simplify the computations for HE, while preserving the accuracy of the model.
We also detail some optimizations that were used to accelerate the runtime.
Finally, we present our results and provide some discussion about our results.

\section*{Notation}
\subsection*{\textbf{Notation for HE}}
Let $N$ be a power-of-two integer and $\mathcal{R} = \mathbb{Z}[x]/\langle x^{N}+1 \rangle$.
For some integer $\ell$, denote $\mathcal{R}_{\ell} = \mathcal{R}/{2^{\ell}}\mathcal{R} = \mathbb{Z}_{2^{\ell}}[x]/\langle x^{N}+1 \rangle$.
We let $\lambda$ be the security parameter where attacks on the cryptosystem require approximately $\Omega (2^{\lambda})$ bit operations.
We use $z \leftarrow \mathcal{D}(Z)$ to represent sampling $z$ from a distribution $\mathcal{D}$ over some set $Z$.
Let $\mathcal{U}$ denote the uniform distribution and $\mathcal{DG}(\sigma^2)$ denote the discrete Gaussian distribution with variance $\sigma^2$.

\subsection*{\textbf{Notation for GWAS}}
The number of samples, covariates and SNPs are denoted as $n, d$ and $k$ respectively.
Matrices are denoted in bold font uppercase letters.
Let the covariates matrix be denoted as $\bfX$ and the SNP matrix as $\bfS$.
The rows of $\bfX$ or $\bfS$ represent the covariates or SNPs from one sample respectively.
We denote the rows as $\bfx_i$.
Vectors are denoted in bold font lowercase letters.
Let the response vector be denoted as $\bfy$.
The vector of weights from the logistic model is denoted as $\bfbeta$ and the corresponding vector of probabilities is denoted as $\bfp$.
The vector of SNP effects is denoted as $\bfs$.
The transpose of a vector $\mathbf{v}$ is denoted as $\mathbf{v}^{\intercal}$.
We let $\POtwo{\cdot}$ denote rounding up to the nearest power-of-two.

\section*{Methods}
\subsection*{\textbf{Homomorphic Encryption}}
HE was first proposed by Rivest \etal \cite{RAD78} more than 40 years ago while the first construction was proposed by Gentry \cite{STOC:Gentry09} only a decade ago.
For this work, we adopt the HE scheme proposed by Cheon \etal \cite{CKKS16}, referred to as CKKS, which enables computation over encrypted approximate numbers.
As GWAS is a statistical function, the CKKS HE scheme is the prime candidate for efficient arithmetic.

Most HE schemes are based on \enquote{noisy} encryptions, which applies some \enquote{small} noise to mask messages in the encryption process.
For HE, a noise budget is determined when the scheme is initialized and computing on ciphertexts depletes this pre-allocated budget.
Once the noise budget is expended, decryption would return incorrect results.
The CKKS scheme \cite{CKKS16} treats encrypted numbers as having some initial precision, with the masking noise just smaller than the precision.
However, subsequent operations on ciphertexts increase the size of noises and reduce the precision of the messages encrypted within.
Thus, decrypted results are approximations of their true value.

The noise budget for the CKKS scheme is initialized with the parameter $L$.
For every multiplication, the noise budget is subtracted by the integer $p$.
The noise budget for a given ciphertext is denoted as $\ell$.
When the message is just encrypted, $\ell = L$.
When $\ell < p$, the noise budget is said to be depleted.

We provide a brief description of the CKKS scheme and highly encourage interested readers to refer to \cite{CKKS16} for the full details.
\begin{itemize}
    \item \textbf{KeyGen($1^{\lambda}$):} \\
    Let $2^{L}$ be the initial ciphertext modulus.
    Let $\mathcal{HWT}(h)$ denote the distribution that chooses a polynomial uniformly from $\mathcal{R}_{2^{L}}$, under the condition that it has exactly $h$ nonzero coefficients.
    Sample a secret $s \leftarrow \mathcal{HWT}(h)$, random $a \leftarrow \mathcal{U}(\mathcal{R}_{2^{L}})$ and error $e \leftarrow \mathcal{DG}(\sigma^2)$.
    Set the secret key as $sk \leftarrow (1, s)$, public key as $pk \leftarrow (b, a) \in  \mathcal{R}_{L}^{2}$ where $b = -a \cdot s + e \pmod{L}$.
    Finally, sample $a' \leftarrow \mathcal{U}(\mathcal{R}_{2^{L}})$, $e' \leftarrow \mathcal{DG}(\sigma^2)$ and set the evaluation key $evk \leftarrow (b', a')$, where $b' = -a' \cdot s + e' + L\cdot s^2 \pmod{2^{2L}}$.

    \item \textbf{Encrypt($pk,m$):} \\
    For $m \in \mathcal{R}$, sample $v \leftarrow \mathcal{U}(\mathcal{R}_{2^{L}})$ and $e_0,e_1 \leftarrow \mathcal{DG}(\sigma^2)$.
    Let $v \cdot pk + (m + e_0, e_1)    \pmod{2^{L}}$ and output $(v,L)$.

    \item \textbf{Decrypt($sk,ct$):} \\
    For $ct = ((c_0, c_1), \ell) \in \mathcal{R}_{\ell}^{2}$, output $c_0 + c_1 \cdot s \pmod{2^{\ell}}$

    \item \textbf{Add($ct_1,ct_2$):} \\
    For $ct_1 = ((c_{0,1}, c_{1,1}), \ell), ct_2 = ((c_{0,2}, c_{1,2}), \ell)$, compute $ (c^{\prime}_0, c^{\prime}_1)\leftarrow = (c_{0,1}, c_{1,1}) + (c_{0,2}, c_{1,2}) \pmod{2^{\ell}}$ and output $(c^{\prime}_0, c^{\prime}_1), \ell)$.

    \item \textbf{Mult($ct_1,ct_2$):} \\
    For ciphertexts $ct_1 = ((c_{0,1}, c_{1,1}), \ell)$ and $ct_2 = ((c_{0,2}, c_{1,2}), \ell)$, let $(d_0, d_1, d_2) = (c_{0,1}c_{0,2}$, $c_{1,1}c_{0,2} + c_{0,1}c_{1,2}, c_{1,1}c_{1,2}) \pmod{2^{\ell}}$.
    Compute $(c^{\prime}_0, c^{\prime}_1) \leftarrow (d_0, d_1) + \lfloor 2^{-L} \cdot d_2 \cdot evk \pmod{2^{\ell}} \rceil$ and output $(c^{\prime}_0, c^{\prime}_1), \ell)$.

    \item \textbf{Rescale($ct,p$):} \\
    For a ciphertext $ct = ((c_0,c_1), \ell)$ and an integer $p \leq \ell$, output $((c^{\prime}_0, c^{\prime}_1), \ell-p)$, where $(c^{\prime}_0, c^{\prime}_1) \leftarrow \lfloor 2^{-p} \cdot (c_0,c_1) \rceil \pmod{2^{\ell-p}} \rceil$.
\end{itemize}

With the CKKS scheme, we are able to encode $N/2$ complex numbers into a single element in its message spaces, $\mathcal{R}$.
This allows us to view a ciphertext as an encrypted array of fixed point numbers.
Let $\phi : \mathbb{C}^{N/2} \rightarrow \mathcal{R}$,
\begin{itemize}
    \item \textbf{Encode($z_1, z_2, \dots, z_{N/2}$):} \\
    Output $m = \phi(z_1, z_2, \dots, z_{N/2})$.

    \item \textbf{Decode($m$):} \\
    Output $(z_1, z_2, \dots, z_{N/2}) = \phi^{-1}(m)$.
\end{itemize}

Informally, $\phi(\cdot)$ maps $(z_1, \ldots, z_{N/2})$ to the vector $(\zeta_j)_{j \in \mathbb{Z}^{\ast}_{N}}$, where $\zeta_j = \lfloor z_{j}\rceil$ and $\zeta_{N-j} = \lfloor\overline{z_{j}}\rceil$ for $1 \leq j \leq N/2$.
This $(\zeta_j)$ is then mapped to an element of $\mathcal{R}$ with the inverse of the canonical embedding map.
$\phi^{-1}(\cdot)$ is straightforward, an element in $\mathcal{R}$ is mapped to a $N$-dimensional complex vector with the complex canonical embedding map and then the relevant entries of the vector is taken to be the vector of messages.

The ability to encode multiple numbers into one ciphertext allows us to reduce the number of ciphertexts used and compute more efficiently.
We refer to each number encoded as a slot of the ciphertext.
This offers a SIMD-like structure where the same computation on all numbers within a ciphertext can be done simultaneously.
This means that adding or multiplying two ciphertexts together would be equivalent to adding or multiplying each slot simultaneously.

The ciphertext of the CKKS scheme can also be transformed into another ciphertext whose slots are a permutation of the original ciphertext.
\begin{itemize}
    \item \textbf{\texttt{Rotate}($ct, r$):}
    Outputs $ct'$ whose slots are rotated to the right by $r$ positions.
\end{itemize}

\subsection*{\textbf{Homomorphic Matrix Operations}}
In this section, we describe our method of encoding matrices with HE.
The batching property of the CKKS scheme allows us to treat ciphertexts as encrypted arrays.
With this, we propose $4$ methods of encoding a matrix with ciphertexts.

\subsubsection*{Column-Packed (CP) Matrices.}
This is our primary method of encoding a matrix.
We encrypt each column of a matrix in one ciphertext and therefore a matrix will be represented by a vector of ciphertexts.
This method of encoding a matrix was suggested by Halevi and Shoup in \cite{HElibAlgo}.

\begin{figure}[h!]
    \centering
    \includegraphics[width=\columnwidth]{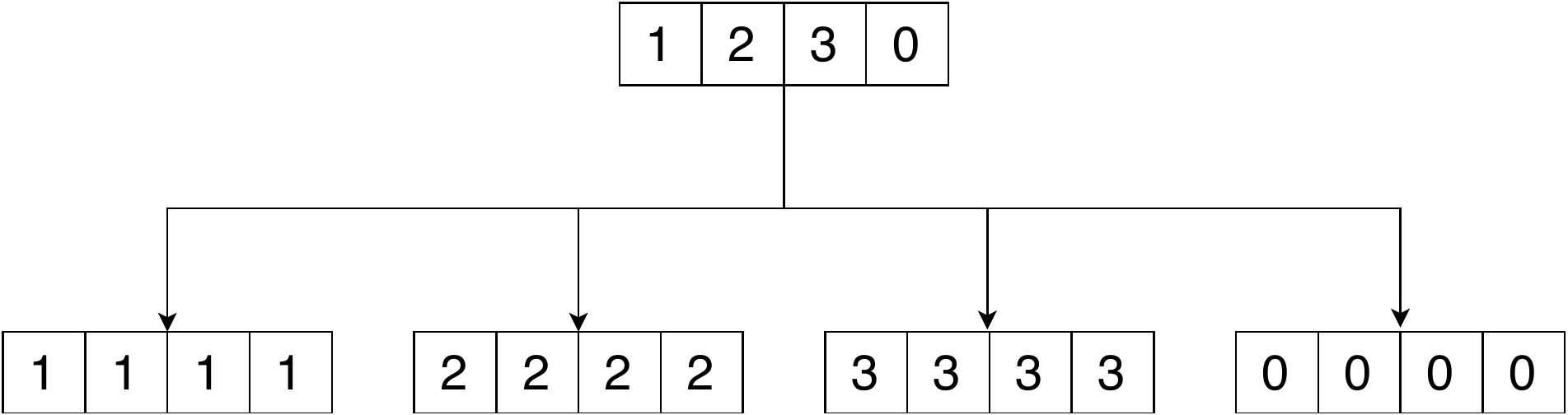}
    \caption{Replicate}
    \label{replicate-figure}
\end{figure}

We require a function, \texttt{Replicate} that takes a vector $\nu$ of size $n$ and returns vectors $\nu_1$, $\nu_2$, $\dots$, $\nu_n$ where $\nu_i$ for $i = 1, \dots , n$, is $\nu[i]$ in all positions.
This is shown in \fref{replicate-figure}.

We describe in \aref{alg:Replicate}, a naive version of \texttt{Replicate}.
The reader is advised to refer to \cite{HElibAlgo} for details on implementing a faster and recursive variant.

\begin{algorithm}
    \caption{\texttt{Replicate}}
    \label{alg:Replicate}

    \SetKwFunction{size}{size}
    \SetKwFunction{Rotate}{Rotate}
    \SetKwData{ct}{ct}
    \SetKwData{result}{result}
    \SetKwData{one}{one}
    \SetKwData{temp}{temp}

    \KwIn{Ciphertext \ct}
    \KwOut{\vbrak{Ciphertext} \result}

    \BlankLine

    \vbrak{int} \one = $[0, 0, \dots , 0]$ \\
    \For{$i = 0$ \KwTo ct.\size}
    {
        \one $[i] = 1$ \\
        Ciphertext \temp = \ct * \one \\
        \For{$j = 0$ \KwTo $\log_{2}(\ct.\size)-1$}
        {
            $\temp \mathrel{{+}{=}} \Rotate(\temp,2^j)$ \\
        }
        $\result[i] = \temp$ \\
    }
\end{algorithm}

We first define matrix-vector multiplication between a CP matrix and a vector in \aref{alg:CPMatVecMult}.
First, we invoke \texttt{Replicate} on the vector.
Next, we multiply each column in the left-hand side matrix with its corresponding $\nu_i$.
Finally, sum up all ciphertexts and this will give the matrix-vector product.

\begin{algorithm}
    \caption{\texttt{CP-MatVecMult}}
    \label{alg:CPMatVecMult}

    \SetKwFunction{Replicate}{Replicate}
    \SetKwFunction{size}{size}
    \SetKwData{colrep}{colrep}
    \SetKwData{result}{result}

    \KwIn{\vbrak{Ciphertext} $A$, Ciphertext $b$}
    \KwOut{Ciphertext \result}

    \BlankLine

    \vbrak{Ciphertext} \colrep $\leftarrow \Replicate(b)$ \\
    \For{$i = 1$ \KwTo \colrep.\size}
    {
        $\result \mathrel{{+}{=}} A(i) * \colrep(i)$ \\
    }
\end{algorithm}

Matrix multiplication between CP matrices is defined as an iterative process over \texttt{CP-MatVecMult} between the left-hand side matrix and the columns of the right-hand side matrix.
This is described in \aref{alg:CP-MatMult}

\begin{algorithm}
    \caption{\texttt{CP-MatMult}}
    \label{alg:CP-MatMult}

    \SetKwFunction{MatVecMult}{CP-MatVecMult}
    \SetKwFunction{size}{size}
    \SetKwData{result}{result}

    \KwIn{\vbrak{Ciphertext} $A$, \vbrak{Ciphertext} $B$}
    \KwOut{\vbrak{Ciphertext} \result}

    \BlankLine

    \For{$i = 1$ \KwTo $B.\size$}
    {
        $\result[i] = \MatVecMult((A, B(i))$ \\
    }
\end{algorithm}

\subsubsection*{Column-Compact-Packed (CCP) Matrices.}
In the case where the entries of a matrix can fit within a single vector, we concatenate its columns and encrypt that in one ciphertext.
For this type of matrix, we are mainly concerned with the function \texttt{colSum} which returns a vector whose entries are the sum of each column.
We present the pseudocode in  \aref{alg:colSum}.
This is achieved by a series of rotations and additions.
However, we do not rotate for all slots of the vector, but rather $\log_2(colSize)$, where $colSize$ is the number of rows in the CCP matrix.
We note here that the final sums are stored in every $colSize$ slots, starting from the first slot.

\begin{algorithm}
    \caption{\texttt{colSum}}
    \label{alg:colSum}

    \SetKwFunction{Rotate}{Rotate}
    \SetKwFunction{colsize}{colsize}
    \SetKwData{result}{result}

    \KwIn{Ciphertext $\nu$}
    \KwOut{Ciphertext \result}

    \BlankLine

    \For{$i = 0$ \KwTo $\log_2(\colsize)-1$}
    {
        $\result \mathrel{{+}{=}} \Rotate(C,2^i)$ \\
    }
\end{algorithm}

\subsubsection*{Row-Packed (RP) Matrices.}
For this encoding, we encrypt rows of a matrix into a ciphertext, representing them with a vector of ciphertexts just like CP matrices.
In this work, we only consider matrix-vector multiplication between an RP matrix and a vector.
Multiplication of an RP matrix by a CP matrix is a lot like naive matrix multiplication.

To compute the multiplication of an RP matrix with a vector, we define the dot product between two vectors encoded in two ciphertexts in \aref{alg:DotProd}.
For that, we first multiply the ciphertexts together, which yields their component-wise products.
Then, we apply rotations to obtain the dot product in every slot of the vector.

\begin{algorithm}
    \caption{\texttt{DotProd}}
    \label{alg:DotProd}

    \SetKwFunction{Rotate}{Rotate}
    \SetKwFunction{size}{size}
    \SetKwData{result}{result}

    \KwIn{Ciphertext $A$, Ciphertext $b$}
    \KwOut{Ciphertext \result}

    \BlankLine

    Ciphertext $C \leftarrow A * B$ \\
    \For{$i = 0$ \KwTo $\log_2(C.\size)-1$}
    {
        $C \mathrel{{+}{=}} \Rotate(C,2^i)$ \\
    }
\end{algorithm}

With \texttt{DotProd}, we apply it over the rows of the RP matrix with the vector, producing several ciphertexts that each contain the dot product between a row and said vector.
Though a series of masks and additions, these separate ciphertexts are combined into the matrix-vector product between an RP matrix and a vector as shown in \aref{alg:RP-MatVecMult}.

\begin{algorithm}
    \caption{\texttt{RP-MatVecMult}}
    \label{alg:RP-MatVecMult}

    \SetKwFunction{DotProd}{DotProd}
    \SetKwFunction{size}{size}
    \SetKwData{result}{result}
    \SetKwData{zero}{zero}

    \KwIn{Ciphertext $A$, Ciphertext $b$}
    \KwOut{Ciphertext \result}

    \BlankLine

    \vbrak{int} \zero = $[0, 0, \dots , 0]$ \\
    \For{$i = 0$ \KwTo $A.\size$}
    {
        $\zero[i] = 1$ \\
        $\result \mathrel{{+}{=}} \DotProd(A(i), b) * \zero$ \\
    }
\end{algorithm}

\subsubsection*{Row-Expanded-Packed (REP) Matrices.}
This method of encoding a matrix is similar to RP matrices, except that each entry is repeated $q$ times for some integer $q$ that is a power of two.
As with RP matrices, REP matrices are represented by vectors of ciphertexts.
By encoding a matrix in this manner, we reduce the number of homomorphic operations when multiplying with other matrices.
For this paper, we only consider matrix products between CP and REP matrices.

\begin{figure}[h!]
    \centering
    \includegraphics[width=0.9\columnwidth]{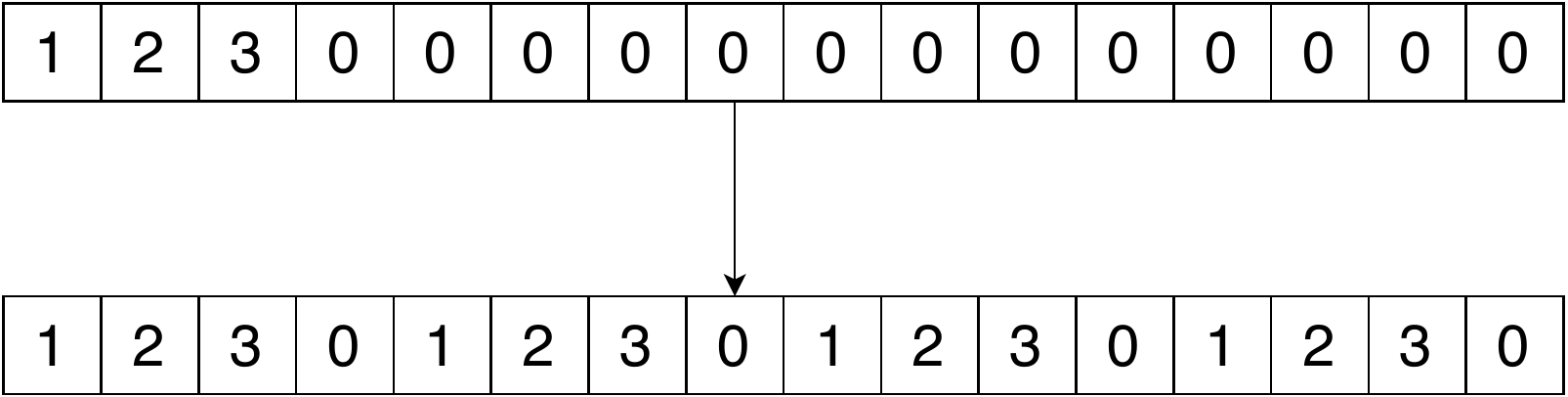}
    \caption{Duplicate}
    \label{duplicate-figure}
\end{figure}

First, we define a function, \texttt{Duplicate} in \aref{alg:Duplicate}.
Suppose that a ciphertext has $k$ filled slots out of $n$, \texttt{Duplicate} fills the remaining slots with repetitions of the $k$ slots.
This is shown in \fref{duplicate-figure}.

This can be realized using simple rotations and additions.

\begin{algorithm}
    \caption{\texttt{Duplicate}}
    \label{alg:Duplicate}

    \SetKwFunction{Rotate}{Rotate}
    \SetKwFunction{colsize}{colsize}
    \SetKwData{result}{result}

    \KwIn{Ciphertext $\nu$}
    \KwOut{Ciphertext \result}

    \BlankLine

    \For{$i = 0$ \KwTo $\log_2(\colsize/\POtwo{k})-1$}
    {
        $\result \mathrel{{+}{=}} \Rotate(C,\POtwo{k})$ \\
    }
\end{algorithm}

To compute matrix products between CP and REP matrices, we first apply \texttt{Duplicate} the columns of the CP matrix.
Then, we multiply each column in the CP matrix with its corresponding row in the REP matrix.
Finally, we sum all the ciphertexts and obtain the product of the matrices in a CCP matrix.
This is shown in \aref{alg:CP-REP-MatMult}.

\begin{algorithm}
    \caption{\texttt{CP-REP-MatMult}}
    \label{alg:CP-REP-MatMult}

    \SetKwFunction{Duplicate}{Duplicate}
    \SetKwFunction{size}{size}
    \SetKwData{result}{result}

    \KwIn{Ciphertext $A$, Ciphertext $B$}
    \KwOut{Ciphertext \result}

    \BlankLine

    \For{$i = 0$ \KwTo $A.\size$}
    {
        $\result \mathrel{{+}{=}} \Duplicate(A(i)) * B(i)$ \\
    }
\end{algorithm}

\subsection*{\textbf{Logistic Regression with Homomorphic Encryption}}
The first step in the GWAS algorithm is to solve a logistic model for its weights $\bfbeta$.
There are several solutions \cite{SecureLR,iDashLR,SEAL-LR,HElib-LR,scalableLR} that solve a logistic model with HE, given that it was one of the challenges in the iDASH 2017 competition.

\subsubsection*{Logistic Regression.}
Logistic regression estimates the parameters of a binary logistic model.
Such models are used to predict the probability of an event occurring given some input features.
These models assume that the logarithm of the odds ratio (log-odds) is a linear combination of the input features.

Let $p$ denote the probability of an event occurring.
The assumption above can be written as
\begin{equation}
    \label{eq:log-odds}
    \log \left( \frac{p}{1-p} \right) =     \beta_0 + \beta_1 x_1 + \cdots + \beta_d x_d.
\end{equation}
Rearranging \cref{eq:log-odds}, we get
\begin{equation}
    \label{eq:sigmoid}
    p(\bfx, \bfbeta) = \frac{1}{1+e^{-\bfbeta^{\intercal} \bfx}}
\end{equation}
where $\bfbeta = (\beta_0, \beta_1, \dots, \beta_d)$ and $\bfx = (1, x_1, \dots, x_d)$.
This is known as the sigmoid function.

Logistic regression estimates the regression coefficients $\bfbeta$ using maximum likelihood estimation (MLE).
This likelihood is given as
\begin{align}
    \label{eq:likelihood}
    \begin{split}
        L(\bfX, \bfbeta) & = \prod_{i=1}^{n} P(y_i|\bfx_i) \\
        & = \prod_{i=1}^{n} p(\bfx_i,\bfbeta)^{y_i} (1- p(\bfx_i,\bfbeta))^{1-y_i}.
    \end{split}
\end{align}
where $\bfx_i$ denotes the rows of the covariates matrix $\bfX$.
Often, MLE is performed with the log-likelihood
\begin{align}
    \begin{split}
        \ell(\bfX, \bfbeta) & = \sum_{i=1}^{n} y_i \log(p(\bfx_i,\bfbeta)) \\
        & \qquad + \sum_{i=1}^{n} (1-y_i) \log(1-p(\bfx_i,\bfbeta)) \label{eq:log-likelihood2}
    \end{split} \\
    \begin{split}
        & = \sum_{i=1}^{n} y_i \log \left( \frac{p(\bfx_i,\bfbeta)}{(1- p(\bfx_i,\bfbeta)} \right) \\
        & \qquad + \sum_{i=1}^{n} \log \left( \frac{1}{e^{\bfbeta^{\intercal} \bfx}+1} \right) \label{eq:log-likelihood3}
    \end{split} \\
    & = \sum_{i=1}^{n} y_i (\bfbeta^{\intercal} \bfx_i) - \sum_{i=1}^{n} \log (e^{\bfbeta^{\intercal} \bfx}+1)
    \label{eq:log-likelihood}
\end{align}

Maximizing \cref{eq:log-likelihood} requires an iterative process.
Our implementation in solving the logistic model applies the Newton-Raphson method \cite{newton-raphson-method}.
This is because the Newton-Raphson method is known to converge quadratically \cite{convergenceNRM} and we wish to solve the model with as little iterations as possible.

The Newton-Raphson method iterates over the following equation
\begin{equation}
    \label{eq:updateBeta}
    \beta^{(t+1)} = \bfbeta^{(t)} - \bfH^{-1}(\bfbeta^{(t)}) \bfg(\bfbeta^{(t)})
\end{equation}
where $\bfg$ and $\bfH$ are given as
\begin{align}
    \label{eq:grad}
    \bfg(\bfbeta) &= \bfXT(\bfy-\bfp(\bfbeta)), \\
    \label{eq:hessian}
    \bfH(\bfbeta) &= \bfXT (\bfW(\bfbeta)) \bfX.
\end{align}

$\bfW(\bfbeta)$ is defined to be a $n$ by $n$ diagonal matrix whose entries are $p_i(1-p_i)$ for $i = 1, \dots , n$.
We remind the reader here that $bfy$ is a $n$ by $1$ binary response vector that contains the truth labels of each individual.
$\bfp(\bfbeta)$ represents the vector of probabilities that is computed for each individual with \cref{eq:sigmoid} using $\bfbeta$ of the particular iteration.

A careful derivation of \cref{eq:log-odds,eq:sigmoid,eq:likelihood,eq:log-likelihood2,eq:log-likelihood3,eq:log-likelihood} can be found in \cite{logisticregression}.

However, there are two non-HE friendly aspects in this algorithm.
Firstly, for each iteration, $\bfH$ is re-computed with the iteration's $\bfbeta$.
This is computationally expensive with homomorphic encryption.
Secondly, the sigmoid function \cref{eq:sigmoid} contains the exponential function, $e^x$ which is not natively supported by HE schemes.
Hence, we approximate the Hessian matrix and the sigmoid function in our implementation.

\subsubsection*{Hessian Matrix Approximation.}
We use an approximation for all Hessian matrices as suggested by B{\"o}hning and Lindsay \cite{Bohning1988}.
They proposed using
\begin{equation}
    \label{eq:hessianApprox}
    \tilde{\bfH} = \frac{1}{4} \bfXT \bfX
\end{equation}
as a lower bound approximation for all Hessian matrices in solving a logistic model with the Newton-Raphson method.
This approximation is also used by Xie \etal~\cite{PrivLogit} in their distributed privacy preserving logistic regression.
We chose to precompute $\left( \bfXT \bfX \right)^{-1}$ with an open source matrix library \textit{Eigen} \cite{eigen}.
We then encrypt $\left( \bfXT \bfX \right)^{-1}$ as an input to the GWAS algorithm.

\subsubsection*{Sigmoid Function Approximation.}
We use the approximation from Kim \etal~\cite{iDashLR} who proposed polynomials of degree $3,5,7$ as approximations of the sigmoid function.
We chose the polynomial of degree $7$:
\begin{multline}
    \label{eq:sigma7}
    \sigma_7(x) = 0.5 - 1.73496 \left( \frac{x}{8} \right) + 4.19407 \left( \frac{x}{8} \right)^3 \\
    - 5.43402 \left( \frac{x}{8} \right)^{5} + 2.50739 \left( \frac{x}{8} \right)^{7}.
\end{multline}

\subsubsection*{Our Algorithm.}
We described our algorithm for Logistic Regression with HE in \aref{alg:LogisticRegression}.
We encrypt $\bfX$ and $\left( \bfXT \bfX \right)^{-1}$ as CP matrices and $\bfy$ as a ciphertext.
We initialize $\bfbeta$ in a ciphertext by encrypting a vector of zeros.
We first compute $\bfX\bfbeta$ with \aref{alg:CPMatVecMult} and apply \cref{eq:sigma7} on to each slot in the ciphertext.
Note here that $\bfX\bfbeta$ is now a vector and is represented by one ciphertext.
Instead of encrypting $\bfXT$, we treat $\bfX$ as $\bfXT$ encrypted as a RP matrix.
We thus invoke RP matrix vector multiplication, \aref{alg:RP-MatVecMult} with $\bfXT$ and $(\bfy - \bfp)$.
Finally, $\bfbeta$ is updated with \cref{eq:updateBeta}.

\begin{algorithm}
    \caption{\texttt{Homomorphic LogReg}}
    \label{alg:LogisticRegression}

    \KwIn{$\bfX$, $\bfy$, $\left( \bfXT \bfX \right)^{-1}$}
    \KwOut{$\bfbeta$}

    \BlankLine

    \For{$i = 1$ \KwTo $\kappa$}
    {
        $\bfp \leftarrow \sigma_7(\bfX\bfbeta)$. \\
        $g \leftarrow \bfXT(\bfy-\bfp)$ \\
        $\tilde{\bfH}^{-1} \leftarrow 4 \left( \bfXT \bfX \right)^{-1}$ \\
        $\bfbeta_{new} \leftarrow \bfbeta - \tilde{\bfH}^{-1}g$\\
        $\bfbeta = \bfbeta_{new}$
    }
\end{algorithm}

In comparison with prior works that perform secure computation of logistic regression with HE \cite{SecureLR,iDashLR,SEAL-LR,HElib-LR,scalableLR}, our method is the first to use the Newton-Raphson method.
Gradient descent was chosen to in maximizing the log-likelihood, \cref{eq:log-likelihood}, in other implementations.

In \cite{SEAL-LR}, a $1$-bit gradient descent method was adopted, with the FV scheme \cite{FV12}.
Bootstrapping is required in this solution.
\cite{SecureLR} employed the CKKS scheme \cite{CKKS16} with gradient descent.
They shared two least squares approximations of the sigmoid function.
The winning solution of iDASH 2018 \cite{iDashLR} used a gradient descent variant - Nesterov Accelerated Gradient and introduced another approximation of the sigmoid function.
\cite{scalableLR} use bootstrapping to achieve logistic regression for datasets larger than any of the solutions published.
A unique solution proposed in \cite{HElib-LR} attempts to approximate a closed form solution for logistic regression.

\subsection*{\textbf{Semi-Parallel GWAS with Homomorphic Encryption}}
The semi-parallel GWAS algorithm proposed by Sikorska \etal \cite{Sikorska2013} rearranges linear model computations and leverages fast matrix operations to achieve some parallelization and thus better performance.
A logistic model is first solved with the covariates matrix.
Let $\bfz$ be a temporary variable
\begin{equation}
    \label{eq:z}
    \bfz = \bfX\bfbeta + \bfW^{-1}(\bfy-\bfp)
\end{equation}
where $\bfbeta$ is the weights of the logistic model, $\bfy$ is the response vector and $\bfp$ is the vector of probabilities from evaluating the sigmoid function \cref{eq:sigmoid} with $\bfbeta$.

The SNP matrix $\bfS$ is then orthogonalized with
\begin{equation}
    \label{eq:Str}
    \bfS^* = \bfS - \bfX \left( \bfXT \bfW \bfX \right)^{-1} \bfXT \bfW \bfS
\end{equation}
and $\bfz$ is orthogonalized with
\begin{equation}
    \label{eq:ztr}
    \bfz^* = \bfz - \bfX \left( \bfXT \bfW \bfX \right)^{-1} \bfXT \bfW \bfz.
\end{equation}

The estimated SNP effect $\bfs$ can then be computed with
\begin{equation}
    \label{eq:SNPeffect}
    \bfb = \frac{(\bfW \bfz^*)^{\intercal} \cdot \bfS^*}{\mathtt{colsum}(\bfW (\bfS^*)^2)}
\end{equation}
and the standard error can be computed with
\begin{equation}
    \label{eq:SNPerror}
    \bferr = \frac{1}{\mathtt{colsum}(\bfW (\bfS^*)^2)}.
\end{equation}
Division here denotes element-wise division between the vectors $(\bfW \bfz)^{\intercal} \cdot \bfS^*$ and $colsum(\bfW (\bfS^*)^2)$.

The main obstacle for HE with the semi-parallel GWAS algorithm is matrix inversion.
General matrix inversion is computationally expensive and inefficient in HE.
This is mainly because integer division, which is used frequently in matrix inversion, cannot be efficiently implemented in HE.
There are two instances where matrix inversion has to be computed.
The first occurs in \cref{eq:z} and the second occurs in the orthogonal transformations \cref{eq:Str,eq:ztr}.
In the following paragraphs, we will describe our method for implementing the semi-parallel GWAS algorithm with HE.
We will also describe some optimizations that reduce memory consumption and accelerate computations to qualify within the competition requirements.

\subsubsection*{Inverse of $\bfW$.}
We exploit the nature of $\bfW$ to compute its inverse with the Newton-Raphson method in HE.
Recall that $\bfW$ is a $n$ by $n$ diagonal matrix whose entries are $p_i(1-p_i)$ for $i = 1, \dots , n$.
Firstly, we represent the diagonal matrix $\bfW$ by a vector $\bfw$ containing the diagonal entries to reduce storage and computational complexity.
Secondly, the inverse of a diagonal matrix is can be obtained by inverting the entries along the main diagonal.
This means that $\bfW^{-1}$ can be computed by inverting the slots of $\bfW$.
The entries of $\bfw$ are given as $p_i(1-p_i)$, where $p_i \in [0,1]$.
We claim an upper bound of $0.25$ on the slots of $\bfw$.
The proof is as follows: the derivative of $p_i(1-p_i)$ is $1-2p_i$ for which $p_i = 0.5$ gives a maximium.
Substituting $p_i = 0.5$ provides the upper bound of $0.25$.

We used this information to set a good initial guess of $3$ in the Newton-Raphson method.
This would reduce the number of iterations needed to obtain an accurate inverse.
We describe this algorithm in \aref{alg:inverseSlots}.

\begin{algorithm}
    \caption{\texttt{inverseSlots}}
    \label{alg:inverseSlots}

    \SetKwFunction{Duplicate}{Duplicate}
    \SetKwFunction{size}{size}
    \SetKwData{w}{$\bfw$}
    \SetKwData{wInv}{$\bfw^{-1}$}
    \SetKwData{guess}{guess}

    \KwIn{$\bfw$}
    \KwOut{$\bfw^{-1}$}

    \BlankLine

    $\guess = [3, 3, \cdots , 3]$ \\
    \For{$i = 1$ \KwTo $3$}
    {
        \wInv = \guess(2 - \w * \guess) \\
        \guess = \wInv \\
    }
\end{algorithm}

\subsubsection*{Modification of Orthogonal Transformations.}
We propose modifications to \cref{eq:Str,eq:ztr} as $\left( \bfXT \bfW \bfX \right)^{-1}$ is too expensive to be computed in the encrypted domain.

We define a placeholder matrix $\bfM$ as
\begin{equation}
    \label{eq:M}
    \bfM = I - \bfX \left( \bfXT \bfX \right)^{-1} \bfXT.
\end{equation}

We proposed a modification, inspired by the Hessian approximation in \cref{eq:hessianApprox}, to the orthogonal transformation of $\bfS$ with
\begin{equation}
    \label{eq:Sprime}
    \begin{split}
        \bfS' & = \bfM \bfS \\
        & = \bfS - \bfX \left( \bfXT \bfX \right)^{-1} \bfXT \bfS
    \end{split}
\end{equation}
and $\bfz$ with
\begin{equation}
    \label{eq:zprime}
    \begin{split}
        \bfz' & =\bfM \bfz \\
        & = \bfz - \bfX \left( \bfXT \bfX \right)^{-1} \bfXT \bfz.
    \end{split}
\end{equation}

The estimated SNP effect is now computed with
\begin{equation}
    \label{eq:estSNPeffect}
    \bfb' = \frac{(\bfW \bfz')^{\intercal} \cdot \bfS'}{\mathtt{colSum}(\bfW (\bfS')^2)}.
\end{equation}
and the standard error is
\begin{equation}
    \label{eq:estSNPerror}
    \bferr' = \frac{1}{\mathtt{colsum}(\bfW (\bfS')^2)}.
\end{equation}

\subsubsection*{Complex Space of CKKS ciphertext}
For our first optimization, we exploit the scheme’s native support for complex numbers to pack two SNPs into a single complex number, putting one SNP in the real part and another in the imaginary part.
This allows us to fit twice as many SNPs in a single ciphertext and cut the runtime by half.

However, $(\bfS_i')^{2}$ in \cref{eq:estSNPeffect} is more difficult to compute with this packing method.
Simply squaring the ciphertext does not yield the correct output as slots now contain complex numbers; for some complex number $z=x+yi$,
\begin{equation}
    \label{eq:z-square}
    (x+yi)^2 = (x^2 - y^2) + 2xyi.
\end{equation}
Instead, we consider multiplying $z$ by its complex conjugate $\overline{z}=x-yi$.
We have
\begin{equation}
    \label{eq:z-zconj}
    (x+yi)(x-yi) = x^2 + y^2
\end{equation}
Extracting the real parts of \cref{eq:z-square} and \cref{eq:z-zconj}, we get
\begin{equation}
    \label{eq:x2y2}
    x^2 = \frac{Re(z\overline{z} + z^2)}{2} \qquad y^2 = \frac{Re(z\overline{z}- z^2)}{2}
\end{equation}
Recall that $\bfS_i'$ is a CCP matrix which is represented by one ciphertext with each slot holding one complex numbers encoding two SNPs.
Thus, we compute $\bfS_i'\bfS_i'$ and $\bfS_i'\overline{\bfS_i'}$.
We assign
\begin{equation}
    \label{eq:S-even}
    \bfS_{even} = \frac{\bfS_i'\overline{\bfS_i'} + \bfS_i'\bfS_i'}{2}
\end{equation}
and
\begin{equation}
    \label{eq:S-odd}
    \bfS_{odd} = \frac{\bfS_i'\overline{\bfS_i'} - \bfS_i'\bfS_i'}{2}
\end{equation}

\subsubsection*{Optimizations with HEAAN}
There are two optimizations that we used with the HEAAN library to reduce the parameters needed and to improve runtime.

For the first optimization, we rescale the ciphertext by a value that is smaller than $p$ after every plaintext multiplication.
This means each plaintext multiplication is now \enquote{cheaper} than a ciphertext multiplication and hence the value of $L$ when initialized can be lowered.

The second optimization would be to perform only power-of-two rotations.
A rotation by $\tau$ slots is a composition of power-of-two rotations in the HEAAN library.
The required power-of-two rotations are the $1$s of the binary decomposition of $\tau$.
Thus, it would be more efficient if we only perform rotations by a power-of-two.
We illustrate this with an example.
A rotation by $245$ slots would require $6$ power-of-two rotations as the binary decomposition of $245$ is $11110101$.
A rotation by $256$ slots would require $1$ power-of-two rotations as the binary decomposition of $256$ is $100000000$.
This reduces the number of rotations in our implementation.

\subsubsection*{Batching SNPs}
As $\bfS$ is too large to be stored in memory when encrypted, we propose to divide $\bfS$ column-wise and process batches of SNPs.
We show how to compute the maximum number of SNPs that can fit within a batch.
Let $\tau$ be the number of SNPs in a batch.
Consider $\bfM\bfS$, a matrix product between a $n$ by $n$ and a $n$ by $\tau$ matrix.
By \aref{alg:CP-REP-MatMult}, the result is a CCP matrix, whose ciphertext has to have enough slots for $n \times \tau$ elements.
For efficiency as described in the previous section, we round the size of each column to the nearest power-of-two and pad the columns with zeroes.
Together with the complex space of the HEAAN ciphertext, the maximum number of SNPs that can be processed as a batch is given as
\begin{equation}
    \label{eq:numSNPinBatch}
    \tau = \frac{2^{N-1}}{2 \times \POtwo{n}}
\end{equation}

\subsubsection*{Smart Cache Module}
We consider the largest matrix in our implementation, $\bfM$ which is a $n$ by $n$ matrix.
There is an instance where $\bfM$ will be stored as CCP matrix (See next section).
This means that the ciphertext would need to have at least ${\POtwo{n}}^2$ slots.
Consequentially, $\log N$ is at least $2 \times {\POtwo{n}}^2$.
This results in a large set of parameters for the HE scheme which translate to a large amount of memory usage.

The next step requires this CCP matrix to be first converted into a CP matrix.
This implies that we need to manage $n$ ciphertexts where $n$ is the number of individuals.
This further increase the memory footprint of the algorithm.

Furthermore, the virtual machine that the iDASH organizers provide only has $16$GB RAM.
As a result, we choose to move ciphertexts to the hard disk when they are not used for computations.

We designed a cache module that exploits the vectorized ciphertext structure of encrypted matrices.
There are 4 threads on the VM provided, of which 2 is used for reading ciphertexts from the disk while 1 is used to write ciphertext into a file.
The last thread is used for computation.
A ciphertext will be pre-fetched into memory before it is needed for computation, replacing a ciphertext that is no longer needed.

\subsubsection*{Our Algorithm}
We give a detailed walkthrough of our modified semi-parallel GWAS algorithm in \aref{alg:SemiParallelGWAS}.

\begin{algorithm}
    \caption{Semi Parallel GWAS}
    \label{alg:SemiParallelGWAS}

    \SetKwFunction{HomLogReg}{HomLogisticRegression}
    \SetKwFunction{inverseSlots}{inverseSlots}
    \SetKwFunction{colSum}{colSum}
    \SetKwFunction{Decrypt}{Decrypt}
    \SetKwFunction{size}{size}
    \SetKwFunction{insert}{insert}

    \SetKwData{WSS}{WSS}
    \SetKwData{WSC}{WSC}
    \SetKwData{denEven}{denEven}
    \SetKwData{denOdd}{denOdd}
    \SetKwData{numerator}{numerator}
    \SetKwData{denominator}{denominator}
    \SetKwData{numeratorEnc}{encNumerator}
    \SetKwData{denominatorEnc}{encDenominator}

    \KwIn{$\bfX$, $\bfbeta$, $\bfy$, $\bfp$, $\left( \bfXT \bfX \right)^{-1}$, $\bfXT$, $\bfS$}
    \KwOut{$\bfb'$}

    \BlankLine
    $\bfbeta$, $\bfp$ $\leftarrow$ \HomLogReg$\left(\bfX, \bfy, \left( \bfXT\bfX \right)^{-1} \right)$ \\
    $\bfw \leftarrow \bfp(\mathbf{1} - \bfp)$ \\
    $\bfw^{-1} \leftarrow \inverseSlots(\bfw)$ \\
    $\bfz \leftarrow \bfX\bfbeta^{(k)} + \bfw^{-1}*(\bfy-\bfp^{(k-1)})$. \\
    \tcc{superscripts for $\bfbeta$ and $\bfp$ indicate the iteration number in \aref{alg:LogisticRegression}}
    $\bfM \leftarrow \mathbf{Id} - \bfX \left( \bfXT \bfX \right)^{-1} \bfXT$ \label{line:computeM} \\
    $\bfz' \leftarrow \bfM \bfz$ \\
    \vbrak{double} \numerator, \denominator \\
    \For{$batch = 1$ \KwTo $\kappa$}
    {
        $\bfS_i' \leftarrow \bfM \bfS_i$ \\
        \numeratorEnc $\leftarrow (\bfw * \bfz')^{\intercal} \cdot \bfS_i'$ \\
        \numerator.\insert(\Decrypt(\numeratorEnc)) \\
        \BlankLine
        $\WSS \leftarrow  (\bfW \bfS_i') \cdot \bfS_i'$ \\
        \BlankLine
        $\WSC \leftarrow  (\bfW \bfS_i') \cdot \overline{\bfS_i'}$ \\
        \BlankLine
        $\denEven \leftarrow 0.5 * (\WSC + \WSS)$ \\
        \BlankLine
        $\denOdd \leftarrow 0.5 * (\WSC - \WSS)$ \\
        \BlankLine
        \For{$i = 0$ \KwTo $\denEven.\size$}
        {
            \denominator.\insert(\Decrypt($\denEven(i)$)) \\
            \denominator.\insert(\Decrypt($\denOdd(i)$)) \\
        }
    }
    \For{$i = 1$ \KwTo $k$}
    {
        $\bfb_i' = \numerator(i) / \denominator(i)$ \\
    }
\end{algorithm}

First, we perform logistic regression with $\bfX$, $\bfy$ and $\left( \bfXT \bfX \right)^{-1}$ as described in \aref{alg:LogisticRegression}.
We use $\bfbeta$ from logistic regression, together with $\bfp$ from the previous iteration to compute $\bfw$ and $\bfz$.

Next, compute the inverse of the slots elements in $\bfw$ with \texttt{inverseSlots}.
Note that $\bfW^{-1}(\bfy-\bfp)$ is equivalent to multiplying the ciphertexts $\bfw^{-1}$ and $(\bfy-\bfp)$.
We then compute $\bfz'$ as given in \cref{eq:zprime}.
At this point, we have $\bfz'$ and $\bfw$ which are both vectors, stored in a ciphertext each.

We construct a temporary variable $\bfM = \mathbf{Id} - \bfX \left( \bfXT \bfX \right)^{-1}$ which is a CP matrix to facilitate computations.
Here, we choose to encrypt $\bfXT$ as a REP matrix.
The reason for encrypting differently is because multiplying a CP matrix by a RP matrix requires the RP matrix to be first converted into REP form.
This process is very inefficient homomorphically and hence we decided to encrypt it directly as a REP matrix.
Thus, the product of $\bfX \left( \bfXT \bfX \right)^{-1}$ with $\bfXT$ is a \texttt{CP-REP-MatMult} as shown in \aref{alg:CP-REP-MatMult}.
At this point, we $\bfM$ is a CCP matrix.
We then convert $\bfM$ into a CP matrix to compute $\bfM\bfS$.

As described earlier, we iterate over partial blocks of the SNP matrix, $\bfS_i$, divided column-wise.
Next, compute its orthogonal transformation $\bfS_i'$.
We remind the reader here again that $\bfM\bfS_i$ is computed with \texttt{CP-REP-MatMult} which produces a CCP matrix, $\bfS'$.
We compute, separately, the numerator, $\bfnum$, and denominator, $\bfden$, of \cref{eq:estSNPeffect} for each $\bfS_i'$.

For $\bfnum$, we multiply $\bfw^{-1}$ and $(\bfz')$ slots-wise and duplicate the slots for as many columns in the CCP matrix $\bfS'$.
The vector-matrix product is now redefined as a ciphertext multiplication, followed by calling \texttt{colSum} over $n$ slots.

For $\bfden$, the computation is similar.
After squaring the slots of the CCP matrix $\bfS'$, we duplicate $\bfw$ and perform a slot-wise multiplication.
\texttt{colSum} of the resulting CCP matrix is exactly the second part of the vector-matrix product for $\bfnum$ - the accumulation sum over every $n$ slots.

We wish to highlight here that as stated in Q$15$ FAQ for the competition, it is acceptable to return $\bfnum$ and $\bfden$ separately \cite{iDashFAQ}.
As such, we decrypt and concatenate all $\bfnum$s and $\bfden$s respectively instead of performing a costly inversion of $\bfden$.
Finally, we divide the two vectors element-wise to obtain the estimated SNP effect, $\bfb'$.

\section*{Results}
We used the provided dataset of $245$ users with $4$ covariates and $10643$ SNPs.

The HE library used is the HEAAN library \cite{HEAAN}, commit id $da3b98$.
The HE parameters used are $\log N = 17$, $\log L = 2440$ and $\log p = 45$.
We observed that the HEAAN context based on these parameters utilizes about $3.5$GB.
The context can be thought of as the base memory needed for HE computations.
Furthermore, we run $\mathbf{Rescale}$ on the output with $p=45$ for ciphertext-ciphertext multiplications and $p=10$ for ciphertext-plaintext multiplications to control noise growth.
This gives us a security level of about $93$ bits based on the LWE estimator provided by Albrecht \etal~\cite{lwe-estimator}.

As described earlier, we require at least ${\POtwo{n}}^2$ slots, where $n=245$.
We chose the minimum number of slots needed, $2^{16}$ slots and set $\log N$ to be $17$.
We are able to process a total of $\tau=512$ SNPs in each batch, based on \cref{eq:numSNPinBatch}.
This gives us a total of $\lceil 10643/512 \rceil = 21$ batches.
We set $\kappa=3$ for the number of iterations in \texttt{HomLogisticRegression}.

We have tabulated the number of sequential homomorphic computations of our modified GWAS algorithm in \tref{table:hom-ops}.

\begin{table}[h!]
    \centering
    \caption{Depth of Homomorphic Operations}
    \label{table:hom-ops}
    \begin{tabular}{lc}
        \toprule
        Homomorphic Operation & No. Successive Operations \\
        \midrule
        Plaintext Multiplication$^*$ & $29$ \\
        Ciphertext Multiplication$^{\dagger}$ & $40$ \\
        Ciphertext Rotation & $256$ \\
        \bottomrule
    \end{tabular}
    \subcaption*
    {
        $^*$ \texttt{Rescale} with $\log p=10$. \\
        $^{\dagger}$ \texttt{Rescale} with $\log p=45$.
    }
\end{table}

These numbers represent the circuit depth of the GWAS algorithm.
We find that a comparison of the number of these computations is a better measure of evaluating a HE program, independent of HE library used.

We report the time taken and memory consumed on two servers: the VM provided by the iDASH organizers and our server.

The machine provided by the iDASH organizers is an Amazon T2 Xlarge or equivalent VM, which has 4 vCPU, 16GB memory, disk size around 200GB \cite{iDashFAQ}.
The results are shown in \tref{table:speed-mem-idash}.

\begin{table}[h!]
    \centering
    \caption{Time Taken and Memory Consumption with iDASH server ($4$ cores) using HEAAN}
    \label{table:speed-mem-idash}
    \begin{tabular}{lcc}
        \toprule
        Process & Time Taken (min) & Memory (GB)\\
        \midrule
        Preprocessing$^*$ & $0.019$ & $0.024$ \\
        Context Generation & $0.65$ & $3.55$ \\
        Encryption & $0.79$ & $0.802628$ \\
        Computations & $717.20$ & $3.98849$ \\
        Decryption & $0.32$ & $0.063$ \\
        \bottomrule
    \end{tabular}
    \subcaption*
    {
        $^*$ Preprocessing time includes file reading, normalizing data and computing $\left( \bfXT \bfX \right)^{-1}$.
    }
\end{table}

For our server, the CPU model used is Intel Xeon Platinum $8170$ CPU at $2.10$GHz with $26$ cores and the OS used is Arch Linux.
The results are shown in \tref{table:speed-mem-heaan}.

\begin{table}[h!]
    \centering
    \caption{Time Taken and Memory Consumption with our server ($22$ cores) using HEAAN}
    \label{table:speed-mem-heaan}
    \begin{tabular}{lcc}
        \toprule
        Process & Time Taken (min) & Memory (GB)\\
        \midrule
        Preprocessing$^*$ & $0.019$ & $0.024$ \\
        Context Generation & $0.43$ & $3.55$ \\
        Encryption & $0.404$ & $0.886795$ \\
        Computations & $203.42$ & $24.1119$ \\
        Decryption & $0.30$ & $0.063$ \\
        \bottomrule
    \end{tabular}
    \subcaption*
    {
        $^*$ Preprocessing time includes file reading, normalizing data and computing $\left( \bfXT \bfX \right)^{-1}$.
    }
\end{table}

We evaluated the accuracy of our results with two methods.
The first method compares the vectors $\bfb$ and $\bfb'$, counting the number of entries that are not equal.
However, since the CKKS scheme introduces some error upon decrypting, we are unable to get any identical entries.
Instead, we opt to count the number of p-values for which our solution differs from the original algorithm by more than some error, $e$.
This is shown in \tref{table:heaan-accuracy}.

\begin{table}[h!]
    \centering
    \caption{HEAAN Accuracy}
    \label{table:heaan-accuracy}
    \begin{tabular}{lcc}
        \toprule
        Error $e$ & No. of Different Entries & HEAAN Accuracy (\%) \\
        \midrule
        $0.1$ & $0$ & $100$ \\
        $0.01$ & $168$ & $98.42$ \\
        $0.005$ & $645$ & $93.94$ \\
        \bottomrule
    \end{tabular}
\end{table}

\begin{figure}[h!]
    \centering
    \includegraphics[width=0.9\columnwidth]{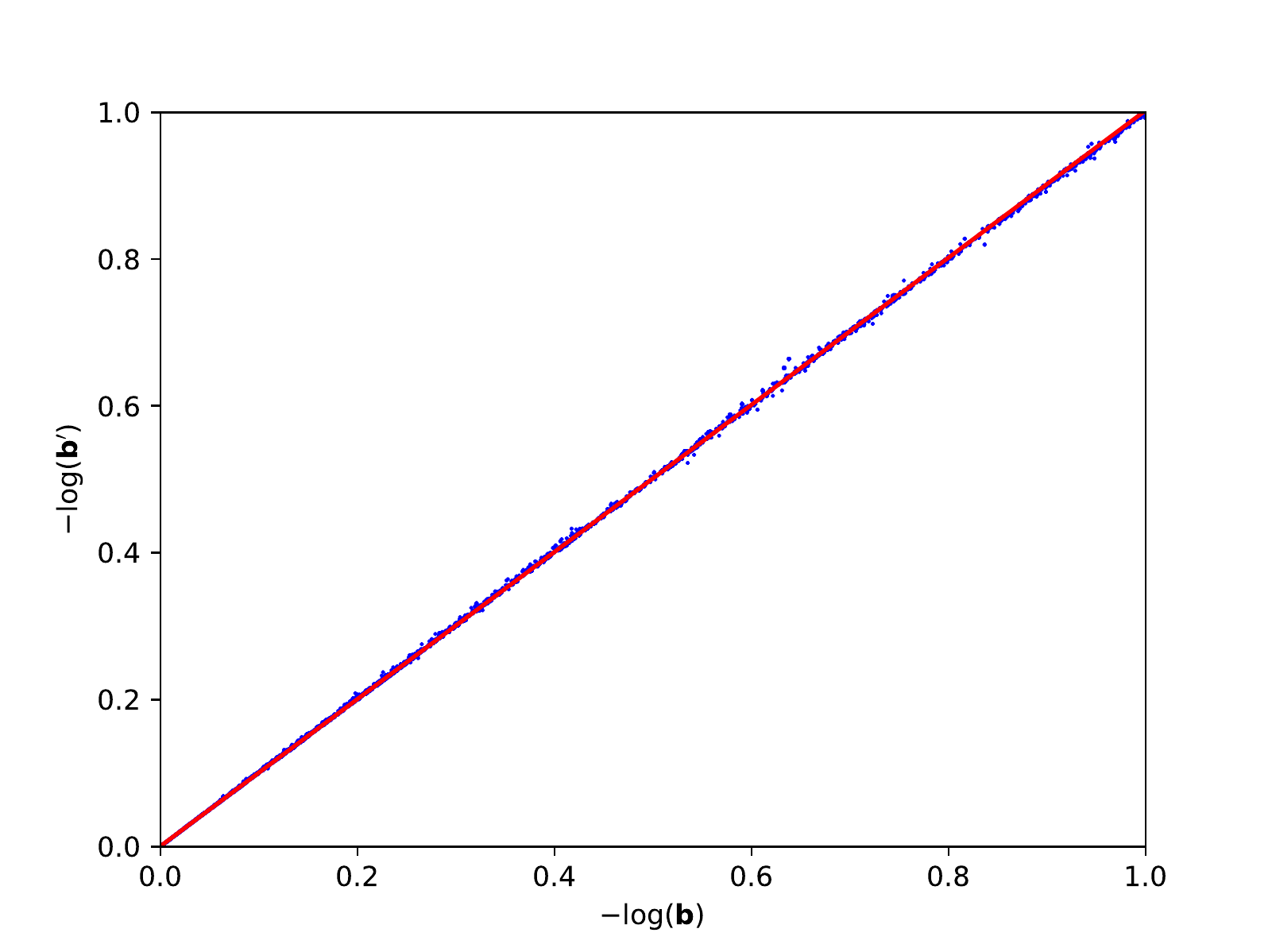}
    \caption{HEAAN Implementation Scatter Plot}
    \label{HEAAN-plot}
\end{figure}

The second method would be to plot a scatter diagram whose x-axis represent $\bfb$ and y-axis represent $\bfb'$.
Ideally, if $\bfb = \bfb'$, the best fit line of the scatter plot should be $y=x$.
We compute the line of best fit with the \texttt{numpy.polyfit} function from python \cite{numpy} and compared against the line $y=x$.
Our HEAAN based solution gives the line $y = 1.002 x + 0.0005317$.
The scatter plot is given in \fref{HEAAN-plot}.

We port our implementation to the SEAL library \cite{SEAL} which recently released a version of the CKKS scheme that
does not require the $2^{2L}$ modulus.
We implemented this with $22$ cores on our machine.
The parameters used are $\log N = 17$, $\log L = 1680$ and $\log p = 50$.
The context generated in this instance is approximately $73.4$GB.
The results of this implementation is given in \tref{table:speed-mem-seal}.

\begin{table}[h!]
    \centering
    \caption{Time Taken and Memory Consumption with our server ($22$ cores) using SEAL}
    \label{table:speed-mem-seal}
    \begin{tabular}{lcc}
        \toprule
        Process & Time Taken (min) & Memory (GB)\\
        \midrule
        Preprocessing$^*$ & $0.020$ & $0.024$ \\
        Context Generation & $12.86$ & $73.4$ \\
        Encryption & $0.20$ & $1.60404$ \\
        Computations & $24.70$ & $38.4843$ \\
        Decryption & $0.31$ & $0.666103$ \\
        \midrule
        Total & $25.21$ & $40.76$ \\
        \bottomrule
    \end{tabular}
    \subcaption*
    {
        $^*$ Preprocessing time includes file reading, normalizing data and computing $\left( \bfXT \bfX \right)^{-1}$.
    }
\end{table}

The accuracy of the SEAL implementation based on the first method is tabulated in \tref{table:seal-accuracy}.

\begin{table}[h!]
    \centering
    \caption{SEAL Accuracy}
    \label{table:seal-accuracy}
    \begin{tabular}{lcc}
        \toprule
        Error $e$ & No. of Different Entries & SEAL Accuracy (\%) \\
        \midrule
        $0.1$ & $127$ & $98.81$\\
        $0.01$ & $4061$ & $61.84$\\
        $0.005$ & $5940$ & $44.19$\\
        \bottomrule
    \end{tabular}
\end{table}

\begin{figure}[h!]
    \centering
    \includegraphics[width=0.9\columnwidth]{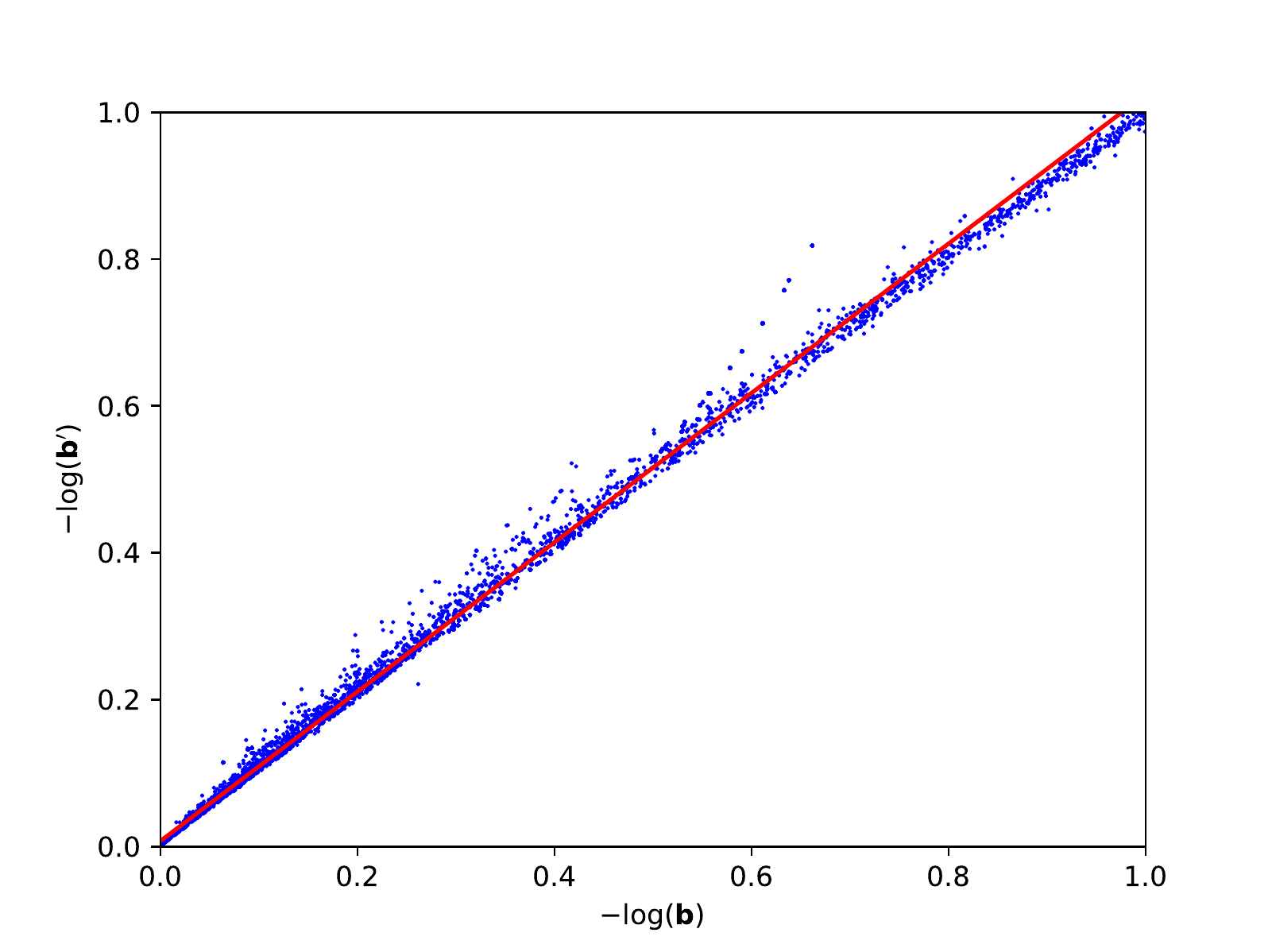}
    \caption{SEAL Implementation Scatter Plot}
    \label{seal-plot}
\end{figure}

Our SEAL based solution gives the line $y = 1.017x + 0.007565$.
The scatter plot for the results is given in \fref{seal-plot}.

\section*{Discussion}
In our submission, we miscalculated the security level, assuming that it fit the $128$-bit requirements while it was actually about $93$ bits.
This is due to the use of the modulus $2^{2L}$ for the evaluation key, which is a quirk of the HEAAN library~\cite{HEAAN}.

There is also a limit of 256 subjects with our implementation, due to our desire to pack the entire test dataset into a single ciphertext.
For a larger number of subjects (up to 512), the matrix $\bfXT$ will need at least $512$ by $512$ slots, which means that $\log N$ has to be at least $19$.

We are aware of the limitations in HEAAN, namely the $2^{2L}$ modulus and slower homomorphic operations.
However, it was the only publicly available HE library based on the CKKS scheme.

We can see that SEAL's implementation of the CKKS scheme is superior in terms of runtime.
This is because SEAL implemented an RNS-variant of the CKKS, which improves the speed of the algorithm by almost $8$ times.
The security level of this implementation based on the LWE estimator is about $230$ bits.

However, we are unable to execute our GWAS algorithm with SEAL using $\kappa=3$.
The set of parameters that supports the depth of the algorithm with $\kappa=3$ appears to be too large and caused our server to run out of memory.
Hence, for the implementation with SEAL, we reduced $\kappa$ to $1$.
This reduces the depth of the algorithm and hence the parameters that were used.
Consequentially, the accuracy of the results has decreased from $98.42\%$ to $61.84\%$.

\section*{Conclusions}
In this paper, we demonstrated an implementation of a semi-parallel GWAS algorithm for encrypted data.
We employed suitable approximations in adapting the semi-parallel GWAS algorithm to be HE-friendly.
Our solution shows that the model trained over encrypted data is comparable to one trained over unencrypted data.
Memory constraints are shown to be of little concern with our implementation of a smart cache, which reduced memory consumption to fit within the limits imposed.
This signifies another milestone for HE, showing that HE is mature enough to tackle more complex algorithms.

\bibliographystyle{bmc-mathphys} 
\bibliography{references}      

\end{document}